\documentclass[aps,preprint,prb]{revtex4}
\usepackage{graphicx}
\usepackage{epsfig}
\usepackage{amsmath}
\usepackage{graphics}
\begin{document}

\author{M. F. Gelin}
\author{D. S. Kosov}

\affiliation{Department of Chemistry and Biochemistry,
 University of Maryland, 
 College Park, 
 MD 20742 }

\title{Velocity dependence of friction and Kramers relaxation rates}

\begin{abstract}

We study the influence of the velocity dependence of friction on 
the escape of a Brownian particle from the deep potential well ($E_{b} \gg k_{B}T$, $E_{b}$ is the barrier height, $k_{B}$ is the Boltzmann constant,  $T$ is the bath temperature).
The bath-induced relaxation is treated within the Rayleigh model (a heavy particle of mass $M$ in the bath of light particles
of mass $m\ll M$) up to the terms of the order of $O(\lambda^{4})$, $\lambda^{2}=m/M\ll1$. The term $\sim 1$ 
is equivalent to the Fokker-Planck dissipative operator, and the term $\sim \lambda^{2}$ 
is responsible for the velocity dependence of friction.
As expected, the correction to the Kramers escape rate in the overdamped limit is proportional to  $\lambda^{2}$ and is small. The corresponding correction in the underdamped limit is proportional to $\lambda^{2}E_{b}/(k_{B}T)$ and is not necessarily small. 
We thus suggest that the effects due to the velocity-dependent friction
may be of considerable importance in determining the rate of escape of
an under- and moderately damped Brownian particle from a deep potential
well, while they are of minor importance for an overdamped particle.
\end{abstract}
\maketitle

\section{Introduction}

Many chemical reactions in a condensed phase can be modeled as an
escape of a Brownian particle from a potential well. The conventional
means for describing this kind of problems are the Fokker-Planck equations
(FPEs), or master equations, or the (generalized) Langevin equations
for the reaction coordinate and the conjugated momentum of the particle
in the external potential. Since the seminal work of Kramers, his
approach has further been refined, and the influences of different
dissipation (collision) mechanisms, \cite{wol78,wol79,wol80,wol80a,sza99}
non-Poissonian collision statistics, \cite{ber99,shu00,shu01} non-Markovian
effects, \cite{zwa59,hyn80,hyn82,nit83} and position-dependent friction
\cite{rob92,pol94,dek98} on the reaction rates have been investigated
(see also reviews \cite{ber88,han90,mel91,pol05}).

In the present paper, we study the effects due to the velocity dependence
of friction. \cite{foot1} Our motivation is as follows. There have
been collected evidences in the literature that relaxations of linear
and angular velocities and kinetic energies in liquids under (highly)
non-equilibrium conditions are reproduced neither by conventional
FPEs with constant frictions nor by master equations with constant
collision frequencies, \cite{gri84,zhu90,rob91,str06,str06a} so that
one has to take into account (linear and/or angular) velocity dependence
of friction. \cite{gri84,zhu90,rob91,str06,str06a,mor74,kam,kos06a}
Indeed, if we consider relaxation under equilibrium conditions, then
all the relevant energies are of the order of $k_{B}T$ ($k_{B}$
being the Boltzmann constant and $T$ being the temperature of the
bath). Putting aside non-Markovian effects,  we can then always introduce
a certain effective velocity-independent friction. This means that  
the effects due to the velocity dependence of friction are normally small
and can safely be neglected. If we consider
relaxation under non-equilibrium conditions, when initial energies of 
the particles are much higher than the typical bath energies $\sim k_{B}T$,
then  the velocity dependence of friction cannot be ignored, in general.

It is not unreasonable to expect that similar effects
may become relevant in the evaluation of the activated
escape rates for underdamped or moderately damped Brownian particles.
If an underdamped particle is trapped in a deep potential
well ($E_{b}\gg k_{B}T$, see Fig. 1) then friction is the rate determining
parameter: fluctuations must supply the particle with the energy $\sim E_{b}\gg k_{B}T$
in order to surmount the barrier. Thus, we have to adequately describe the Brownian particle
relaxation not only within the thermal energy interval $\sim k_{B}T$ but also for much higher energies 
$\sim E_{b}$. In such a case the constant friction approximation might be an oversimplification, 
so that the velocity dependence 
of friction may become crucial. On the contrary, the effect of the
velocity-dependent friction on the the overdamped Brownian particle
is expected to be minor, since dissipation manifests itself through
the diffusion coefficient which equals to the integral relaxation
time of the \textit{equilibrium} velocity correlation function, which is determined by 
the typical energies $\sim k_{B}T$. The
present paper is aimed at proving and elucidating the above qualitative
expectations.

\section{Escape rates in Rayleigh model}

All the models studied so far within the framework of the theory of
activated rate processes treat dissipation either via the FPE with
velocity-independent friction or or via master equations with velocity-independent
collision frequency. \cite{wol78,wol79,wol80,wol80a,sza99,ber99,shu00,shu01}
In order to take into account velocity-sensitive dissipation, we must
start either from the FPE with velocity-dependent friction \cite{kos06a}
or from master equations with velocity-dependent collision rates. \cite{kam,hoa71,rid72}
To avoid introducing phenomenological dissipation models, we use the
Rayleigh model of the nonlinear Brownian motion, which describes a one-dimensional (1D)
relaxation of a heavy particle of mass $M$ in the bath of light particles
of mass $m\ll M$. Within this model, collisional relaxation of the
heavy particle is described by the master equation with the velocity-dependent
collision frequency. \cite{kam,hoa71} The model possesses a small
parameter $\lambda^{2}=m/M\ll1$. 
Then, according to van Kampen \cite{kam},
we can run a systematic expansion of this master equation, which allows
us to construct the dissipation operator to any desirable order in
$\lambda$. Up to the terms of the order of $O(\lambda^{4})$, our
starting equation reads \cite{kam65,ply05}

\begin{equation}
\partial_{t}\rho(x,v,t)=\left\{ -v\partial_{x}+M^{-1}U'(x)\partial_{v}+\xi(L_{1}+\lambda^{2}L_{2})\right\} \rho(x,v,t).\label{FPE0}\end{equation}
Here $\rho(x,v,t)$ is the probability density function in the position
($x$) -- velocity ($v$) phase space, $U(x)$ is an external potential
which is schematically depicted in Fig. 1, \begin{equation}
\xi=\frac{8\eta}{\sqrt{2\pi}}\lambda^{2}\sigma^{-1/2}\label{fr}\end{equation}
 is the friction, $\eta$ is the number of the bath particles per
unit length, \begin{equation}
\sigma=\beta M,\,\,\,\beta=\frac{1}{k_{B}T}.\label{P1}\end{equation}
The dissipation operator in Eq. (\ref{FPE0}) consists of two parts.
\begin{equation}
L_{1}=\partial_{v}v+\frac{1}{\sigma}\partial_{v}^{2}\label{L1}\end{equation}
is the standard Fokker-Planck operator, while the next-order correction
to it is explicitly written as \begin{equation}
L_{2}=-\partial_{v}v+\frac{\sigma}{6}\partial_{v}v^{3}-\frac{2}{\sigma}\partial_{v}^{2}+\frac{3}{2}\partial_{v}^{2}v^{2}+\frac{8}{3\sigma}\partial_{v}^{3}v+\frac{4}{3\sigma^{2}}\partial_{v}^{4}.\label{L2}\end{equation}

After doing some algebra, we can rewrite
Eq. (\ref{FPE0}) in the following compact form: 

\begin{equation}
\partial_{t}\rho(x,v,t)=\left\{ -v\partial_{x}+M^{-1}U'(x)\partial_{v}+\xi\partial_{v}\rho_{B}(v)\hat{S}(v)\partial_{v}\rho_{B}^{-1}(v)\right\} \rho(x,v,t).\label{FPE}\end{equation}
 Here \begin{equation}
\hat{S}(v)=1+\lambda^{2}\delta+\lambda^{2}p(\partial_{v}+(q-\sigma)v)(\partial_{v}-qv)\label{S}\end{equation}
is the core of the dissipation operator, and\begin{equation}
\rho_{B}(v)=\sqrt{\sigma/(2\pi)}\exp\{-\sigma v^{2}/2\}\label{Bol1}\end{equation}
 is the equilibrium Boltzmann distribution, and \begin{equation}
p=\frac{4}{3\sigma^{2}},\,\,\, q=\sigma\frac{2\pm\sqrt{2}}{4},\,\,\,\delta=\pm\frac{\sqrt{2}}{3\sigma}.\label{p}\end{equation}
 The standard Kramers FPE is recovered in the limit $\lambda^{2}\rightarrow0$,
$\hat{S}(v)\rightarrow1$. Eq. (\ref{FPE}) can be solved numerically
for any values of parameters (\ref{p}). To get a better understanding
of the effect of velocity-dependent friction on the escape rates,
we analyze the problem analytically. 

Applying the projector operator technique, we can reduce 2D Eq. (\ref{FPE})
to the corresponding 1D equations in the limit of the high ($\xi\gg\Omega$)
and weak ($\xi\ll\Omega$) friction, $\Omega$ being a characteristic
well frequency. In the former case, the Brownian particle's velocity
is a fast variable. We can thus introduce the projection operators
\[
P=\rho_{B}(v)\int dv,\,\,\, Q=1-P,\]
and arrive at the Smoluchowski diffusion equation for the position-dependent
probability density:\begin{equation}
\partial_{t}\rho(x,t)=D\partial_{x}\left\{ \beta U'(x)+\partial_{x}\right\} \rho(x,t).\label{Sm}\end{equation}
Here the diffusion coefficient, $D$, equals to the integral relaxation
time of the equilibrium velocity correlation function in the overdamped limit, $\int_{0}^{\infty}dt\left\langle vv(t)\right\rangle $.
Within the present model this correlation function has been evaluated
by Hynes, \cite{hyn75} so that \begin{equation}
D=\frac{1}{\xi\beta}\{1+\frac{\lambda^{2}}{18}\}+O(\lambda^{4})\label{Dlia}\end{equation}
 and the Kramers rate (in the limit of low temperature and/or high barrier height, $\beta E_{b}\gg1$)
reads \begin{equation}
k=D\frac{\omega_{0}\omega_{b}}{2\pi}\exp\{-\beta E_{b}\}+O(\lambda^{4}).\label{Kr_h}\end{equation}
The frequencies $\omega_{0}$ and $\omega_{b}$ determine the shape
of the potential linearized in the vicinity of the bottom of the well
and at the barrier,\begin{equation}
U(x)\approx\omega_{0}(x-x_{0})^{2}/2 \,\,\, \mathrm{and} \,\,\, U(x)\approx E_{b}-\omega_{b}(x-x_{b})^{2}/2,\label{Ho}\end{equation}
respectively, see Fig. 1. As expected, the velocity-dependent friction increases the
diffusion coefficient, \cite{gri84,hyn75} but the effect is of the
order of $\lambda^{2}$ and is therefore small.

The underdamped limit ($\xi\ll\Omega$) is much more interesting.
We can either switch to the action ($I$) -- angle ($\phi$) variables
and introduce the projectors \[
P=\frac{1}{2\pi}\int_{0}^{2\pi}d\phi,\,\,\, Q=1-P,\]
or change to the energy variable, $E$. In any case, we can start
from Eq. (\ref{FPE}) and arrive at the energy FPE

\begin{equation}
\partial_{t}\rho(E,t)=\xi\partial_{E}\hat{D}(E)\rho_{eq}(E)\partial_{E}\rho_{eq}^{-1}(E)\rho(E,t).\label{FPEE}\end{equation}
The core of the energy FPE operator (\ref{FPE}) is
explicitly written as \begin{equation}
\hat{D}(E)=D_{0}(E)+D_{1}(E)\partial_{E}+D_{2}(E)\partial_{E}^{2};\label{DE}\end{equation}
\begin{equation}
D_{0}(E)=\frac{2}{\beta}\{ b_{1}(E)+\lambda^{2}(\beta b_{3}(E)/3-b_{1}(E))\},\label{DE0}\end{equation}
\begin{equation}
D_{1}(E)=\frac{8\lambda^{2}}{3\beta}\{3b_{1}(E)/\beta-2b_{3}(E)\},\label{DE1}\end{equation}
\begin{equation}
D_{2}(E)=\frac{16\lambda^{2}}{3\beta^{2}}b_{3}(E).\label{DE2}\end{equation}
The equilibrium distribution reads \begin{equation}
\rho_{eq}(E)=Z_{E}^{-1}a_{-1}(E)\exp\{-\beta E\},\,\,\, Z_{E}=\int dEa_{-1}(E)\exp\{-\beta E\},\label{roE}\end{equation}
and we have introduced the quantities \begin{equation}
a_{n}(E)=\int dx(E-U(x))^{n/2},\,\, b_{n}(E)=a_{n}(E)/a_{-1}(E)\label{ab}\end{equation}
(integration is presumed over all $x$ for which $E>U(x)$). The standard
Kramers energy-diffusion FPE \cite{hyn82,nit83,zwa} is recovered
in the limit $\lambda^{2}\rightarrow0$. Eq. (\ref{FPEE}) can be
used to run perturbative series in $\lambda^{2}$ to calculate the
mean first passage time, $\tau(E)$, the inverse of which yields the
relaxation rate, $k$ (see Appendix). If we again assume that $\beta E_{b}\gg1$,
we get then
\begin{equation}
k=k_{Kr}\left[1+\lambda^{2}\frac{\beta a_{3}(E_{b})}{3a_{1}(E_{b})}\right]+O(\lambda^{4}).\label{kSmall}\end{equation}
This is the main result of the present paper. If we neglect the $\sim\lambda^{2}$
correction, then we recover the Kramers expression for the rate in
the underdamped limit \cite{hyn82,nit83,zwa} 
\begin{equation}
k_{Kr}=2\xi\beta\exp\{-\beta E_{b}\}\frac{a_{1}(E_{b})}{a_{-1}(0)}.\label{Kram}\end{equation}
On the other hand,
the expression in the square brackets in Eq. (\ref{kSmall}) can easily be evaluated. For
the harmonic oscillator (\ref{Ho}), for example, it equals $1+\lambda^{2}\beta E_{b}/4$.
For the Morse oscillator it yields $1+\lambda^{2}\beta E_{b}/6$.
Therefore, the actual small parameter of the problem in the underdamped
limit is $\lambda^{2}\beta E_{b}$ rather than $\lambda^{2}$. Since
Eq. (\ref{kSmall}) has been derived in the limit $\beta E_{b}\gg1$,
the product $\lambda^{2}\beta E_{b}$ can be $\sim1$ or higher even
for $\lambda^{2}\ll1$. In this case, evidently, the perturbative
expansion of the dissipative operator in $\lambda^{2}$ may break
down, and the description within the Rayleigh master equation \cite{kam,hoa71}
might be necessary. 

We did not consider the influence of non-Markovian effects on the escape
rates. These effects normally reduce molecular friction and therefore
decrease the escape rates in the underdamped limit. Their influence
is just the opposite to that of the velocity-dependent friction. However,
as has been pointed out in, \cite{hyn82,nit83} the non-Markovian
effects are of minor importance in the case of a deep well, $\beta E_{b}\gg1$.
It is in this latter case we predict the velocity dependence of friction
can give rise to a dramatic increase of the escape rate, $\sim\lambda^{2}(E_{b}\beta)^{1}$,
despite its influence on the velocity relaxation under equilibrium
conditions is small, $\sim\lambda^{2}(E_{b}\beta)^{0}\ll1$.

\section{Conclusions }

The main result of our paper can be formulated as follows: the effects due to the velocity-dependent
friction \cite{foot1} may be of considerable importance in determining the 
rate of escape of an under- and moderately damped Brownian particle
from a deep potential well. Similar effects due to the (possible)
energy dependence of vibrational relaxation rates might also be significant
in describing unimolecular reactions under non-equilibrium conditions,
like those investigated in. \cite{pol97,pol98}

\begin{acknowledgments}
This work was partially supported by the American Chemical Society
Petroleum Research Fund (44481-G6). 
\end{acknowledgments}

\appendix
\section{Calculation of the first passage time in the underdamped limit}

The first passage time in the underdamped limit, $\tau(E)$, can be
evaluated through the equation \cite{zwa}

\begin{equation}
-1=\rho_{eq}^{-1}(E)\partial_{E}\hat{D}(E)\rho_{eq}(E)\partial_{E}\tau(E)\label{FPEE1}\end{equation}
obeying the boundary conditions $\tau'(0)=0$, $\tau(E_{b})=0$. The
adjoined energy FPE operator, $\rho_{eq}^{-1}(E)\partial_{E}\hat{D}(E)\rho_{eq}(E)\partial_{E}$,
is explicitly defined via Eqs. (\ref{DE})-(\ref{ab}). \cite{foot2} In order to
solve Eq. (\ref{FPEE1}) perturbatively, note that $D_{0}(E)$ possesses
contributions both $\sim1$ and $\sim\lambda^{2}$, while $D_{1}(E)$
and $D_{2}(E)$ are both $\sim\lambda^{2}$. Thus we get \begin{equation}
\tau(E)=\int_{E}^{E_{b}}dE'\frac{\rho_{eq}^{-1}(E')}{D_{0}(E')}\left\{ 1-\hat{Y}(E')\right\} \int_{0}^{E'}dE''\rho_{eq}(E'')+O(\lambda^{4}),\label{Tau1}\end{equation}
\begin{equation}
\hat{Y}(E)=\rho_{eq}(E)\left\{ D_{1}(E)\partial_{E}+D_{2}(E)\partial_{E}^{2}\right\} \frac{\rho_{eq}^{-1}(E)}{D_{0}(E)}.\label{Y}\end{equation}
We further assume that the temperature is low, $\beta E_{b}\gg1$.
Then, taking into account the explicit form of $D_{1}(E)$ and $D_{2}(E)$,
we see that the leading contributions stemming from $\hat{Y}(E)$, which are
of the order of $\lambda^{2}(E_{b}\beta)^{0}$, cancel each other.
Thus, $\hat{Y}(E)$ yields contributions $\sim\lambda^{2}(E_{b}\beta)^{-1}$ , which  can be neglected
as compared with those $\sim\lambda^{2}(E_{b}\beta)^{0}$ and
$\sim\lambda^{2}(E_{b}\beta)^{1}$. Then, omitting $\hat{Y}(E)$ in
Eq. (\ref{Tau1}) and evaluating integrals to the leading order, we
arrive at the expression\begin{equation}
\tau(E)=\frac{1}{\xi\beta^{2}}a_{-1}(E_{b})\exp\{\beta E_{b}\}\frac{a_{-1}(0)}{D_{0}(E_{b})}+O(\lambda^{4}).\label{Tau2}\end{equation}
Finally, retaining only the leading contribution $\sim\lambda^{2}(E_{b}\beta)^{1}$
into $D_{0}(E_{b})$ (\ref{DE0}) and inverting Eq. (\ref{Tau2})
one gets Eq. (\ref{kSmall}).

\clearpage
\begin{figure}
\includegraphics[keepaspectratio,totalheight=12cm]{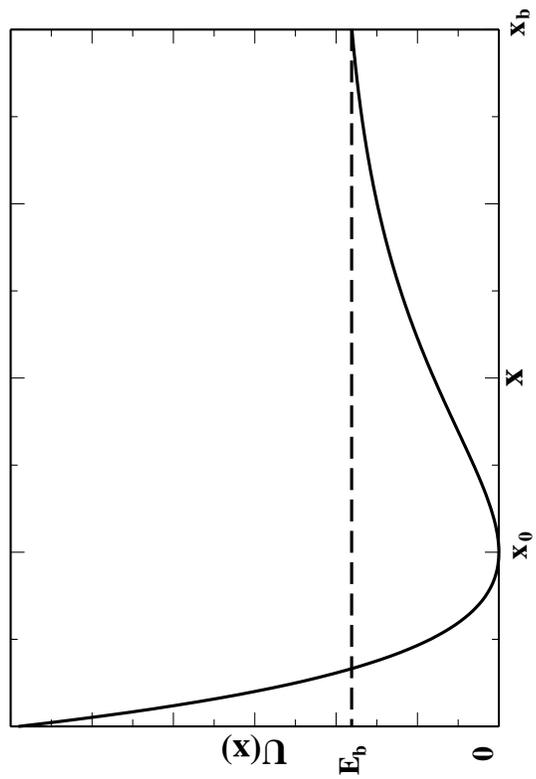}
\caption{Sketch of the potential energy surface.}
\end{figure}


\begin{thebibliography}{20}
\bibitem{foot1}These effects can also be regarded as nonlinear and non-Gaussian,
since the standard harmonic oscillator bath generates the generalized
Langevin equations and FPEs with velocity-independent friction.
\bibitem{foot2} The energy-FPE operators in Eqs. (\ref{FPEE}) and (\ref{FPEE1}) can be regarded as the forward and 
backward Kolmogorov operators, respectively.
\end{thebibliography}
\end{document}